\documentclass[journal]{IEEEtran}
\usepackage{cite}     
\usepackage{graphicx}

\usepackage{amsmath}

\begin{document}

\title{Calibration of a Highly Granular Hadronic Calorimeter with SiPM Readout}

\author{{Frank Simon \\ {\it{on behalf of the CALICE Collaboration}}}
\thanks{Manuscript submitted on November 14, 2008}
\thanks{F.~Simon is with the Max-Planck-Institut f\"ur Physik, Munich, Germany and with the Excellence Cluster Universe, Technical University Munich, Germany.({\it email: frank.simon@universe-cluster.de}).}
}

\maketitle

\begin{abstract}
A highly granular hadronic calorimeter (HCAL) based on scintillator tiles with individual readout by silicon photomultipliers (SiPMs) has been constructed by the CALICE collaboration and has been tested extensively in particle beams at CERN. The 7608 SiPMs coupled to scintillator tiles in the approximately 1 cubic meter large calorimeter allow large sample studies of behavior of these devices under varying voltage and temperature as well as their saturation behavior. We also present detailed studies of the temperature dependence of the calorimeter signal in muon and hadron beams. The calibration of the full calorimeter using muons is discussed. In addition, a novel method for calibration and detector studies using minimum-ionizing track segments in hadronic showers, is explored. 
\end{abstract}


\maketitle
\thispagestyle{empty}

\section{Introduction}

The goal of the CALICE experimental program is to establish novel technologies for calorimetry in future collider
experiments and to record electromagnetic and hadronic shower data with unprecedented three dimensional spatial resolution for the
validation of simulation codes and for the test and development of reconstruction algorithms. Such highly granular calorimeters are necessary to achieve an unprecedented jet energy resolution at the International Linear Collider \cite{:2007sg} using particle flow algorithms \cite{Thomson:2008zz}.

\begin{figure}
\centering
\includegraphics[width=0.48\textwidth]{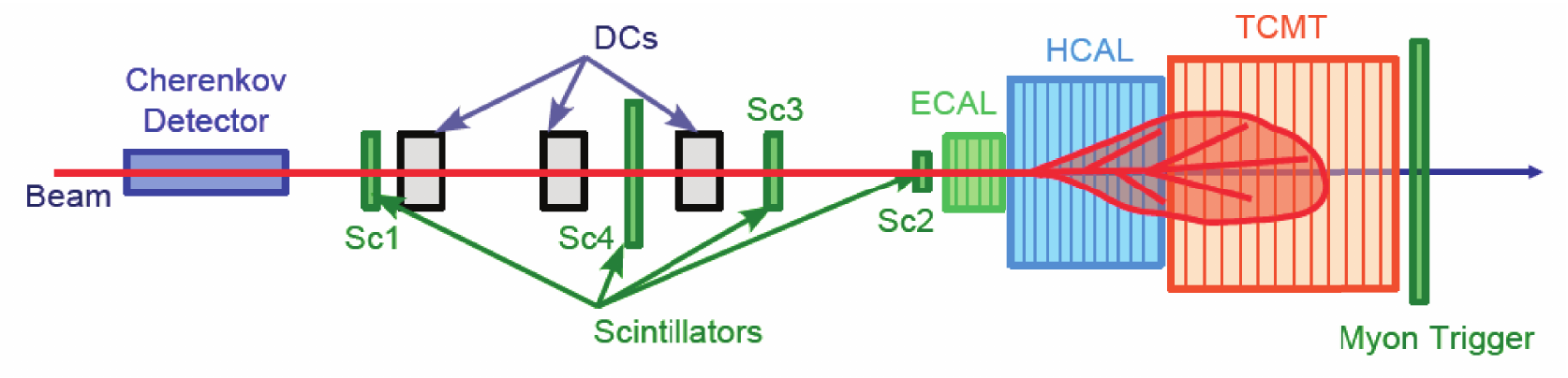}
\caption{Schematic of the CALICE experimental setup at
CERN, with electromagnetic and hadronic calorimetry as
well as a tail catcher and muon tracker downstream of the
calorimeters.}
\label{fig:CALICESetup}
\end{figure}

The CALICE test beam setup \cite{Eigen:2006eq} consists of a silicon-tungsten electromagnetic calorimeter (ECAL), an analog scintillator-steel hadron calorimeter (AHCAL) and tail catcher/muon tracker (TCMT), the latter two both with individual cell readout by silicon photomultipliers (SiPMs) \cite{Bondarenko:2000in}. This setup has been tested extensively in electron, muon and hadron beams at CERN and at the Meson Test Beam Facility at Fermilab. Figure \ref{fig:CALICESetup}  shows the schematic setup of the CALICE detectors in the CERN H6 test beam area, where data was taken in 2006 and 2007.

The AHCAL is a 38 layer sampling calorimeter with 1.6 cm thick steel absorber plates and scintillator layers build out of
individual tiles housed in steel cassettes with a wall thickness of 2 mm, resulting in a total absorber thickness of 2 cm per layer. The lateral dimensions are roughly 1$\times$1 m$^2$, the total thickness amounts to 4.5 nuclear interaction lengths. The first 30 layers of the calorimeter have a high granular core of 10$\times$10 tiles with a tile size of 30$\times$30 mm$^2$, an outer core composed of 60x60 mm$^2$ tiles, and border tiles with a size of 120$\times$120 mm$^2$. The last 8 layers use only 60$\times$60 mm2 tiles in the core, and the large border tiles. In total, this amounts to 7608 channels. The light in each scintillator cell is collected by a wavelength shifting fiber, which is coupled to the SiPM. The SiPMs, produced by the MEPhI/PULSAR group \cite{Bondarenko:2000in}, have a photo-sensitive area of 1.1$\times$1.1 mm$^2$ containing 1156 pixels with a size of 32$\times$32 $\mu$m$^2$.

In this paper, we discuss the calibration of the AHCAL using a built-in LED calibration system as well as a cell-by-cell calibration using muon data. Here, the large available data set of a detector with almost almost 8000 SiPMs connected to scintillator tiles permits large sample studies of various properties of the system, such as the temperature dependence of the response of the system. The hadronic data set is used to explore the possibilities of calibrating without muons and to study the temperature response over a wide range of temperatures.

\section{Built-in calibration systems}

\begin{figure}
\centering
\includegraphics[width=0.495\textwidth]{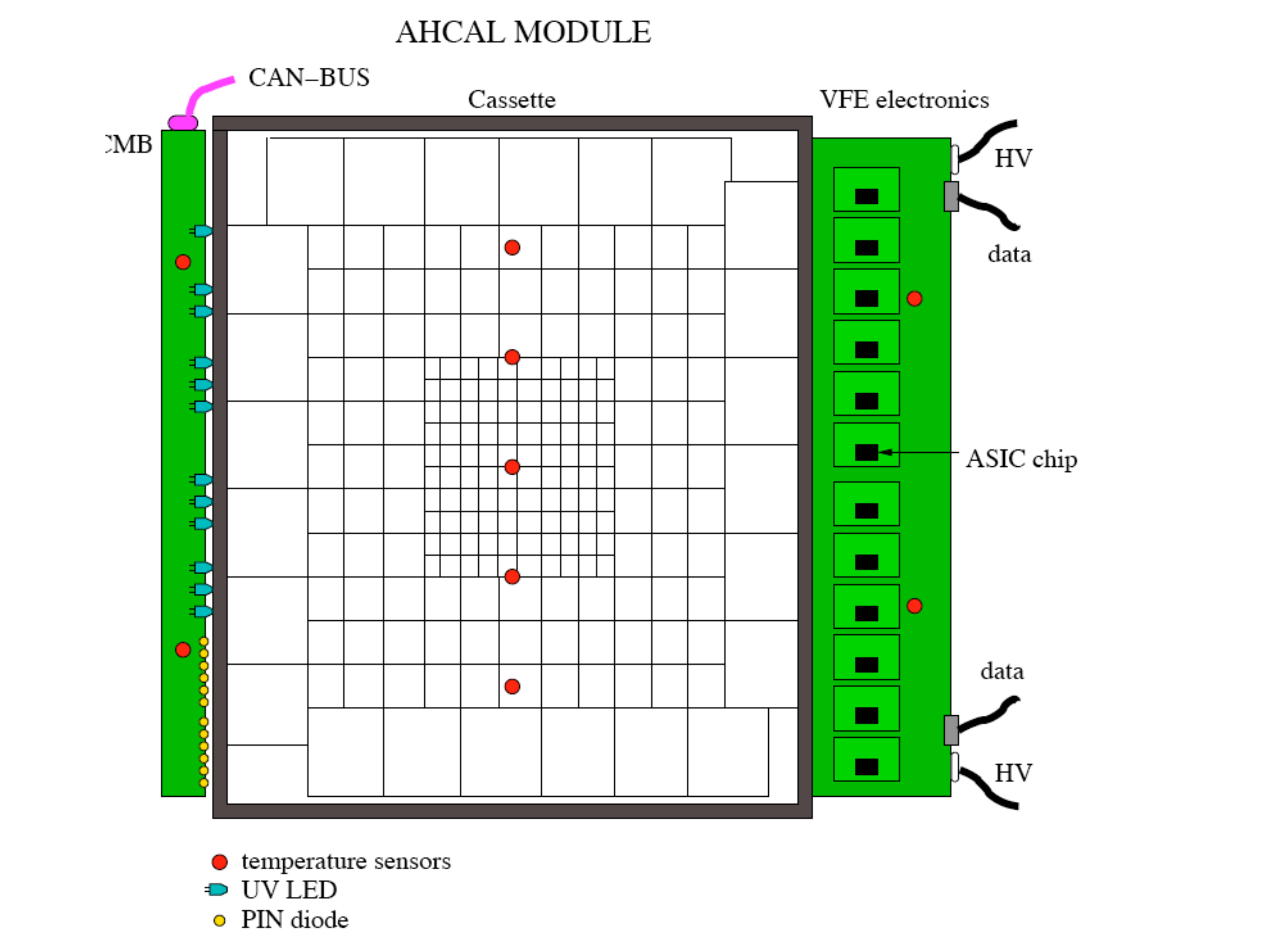}
\caption{Layout of one readout module of the AHCAL. Shown are the scintillator cells, with a high granular core in the center of the layer, the position of five temperature sensors installed in each layer, as well as the LED system and the front-end electronics.}
\label{fig:ModuleLayout}
\end{figure}

The readout modules of the HCAL have extensive monitoring and calibration capabilities built in. Each module is equipped with five temperature sensors to monitor environmental conditions, placed at different vertical positions in the middle of the module.  With this system, the temperature of each cell in the detector can be determined with good accuracy, allowing detailed studies of the temperature behavior of the whole detector system. A LED system with UV LEDs, coupled to each of the cells by clear fibers, and equipped with PIN diodes to monitor the LED light intensity, is available. Figure \ref{fig:ModuleLayout} shows the layout of a readout layer for the first 30 layers of the calorimeter, which include the high granular core with 30$\times$30 mm$^2$ scintillator cells.

\begin{figure}
\centering
\includegraphics[width=0.48\textwidth]{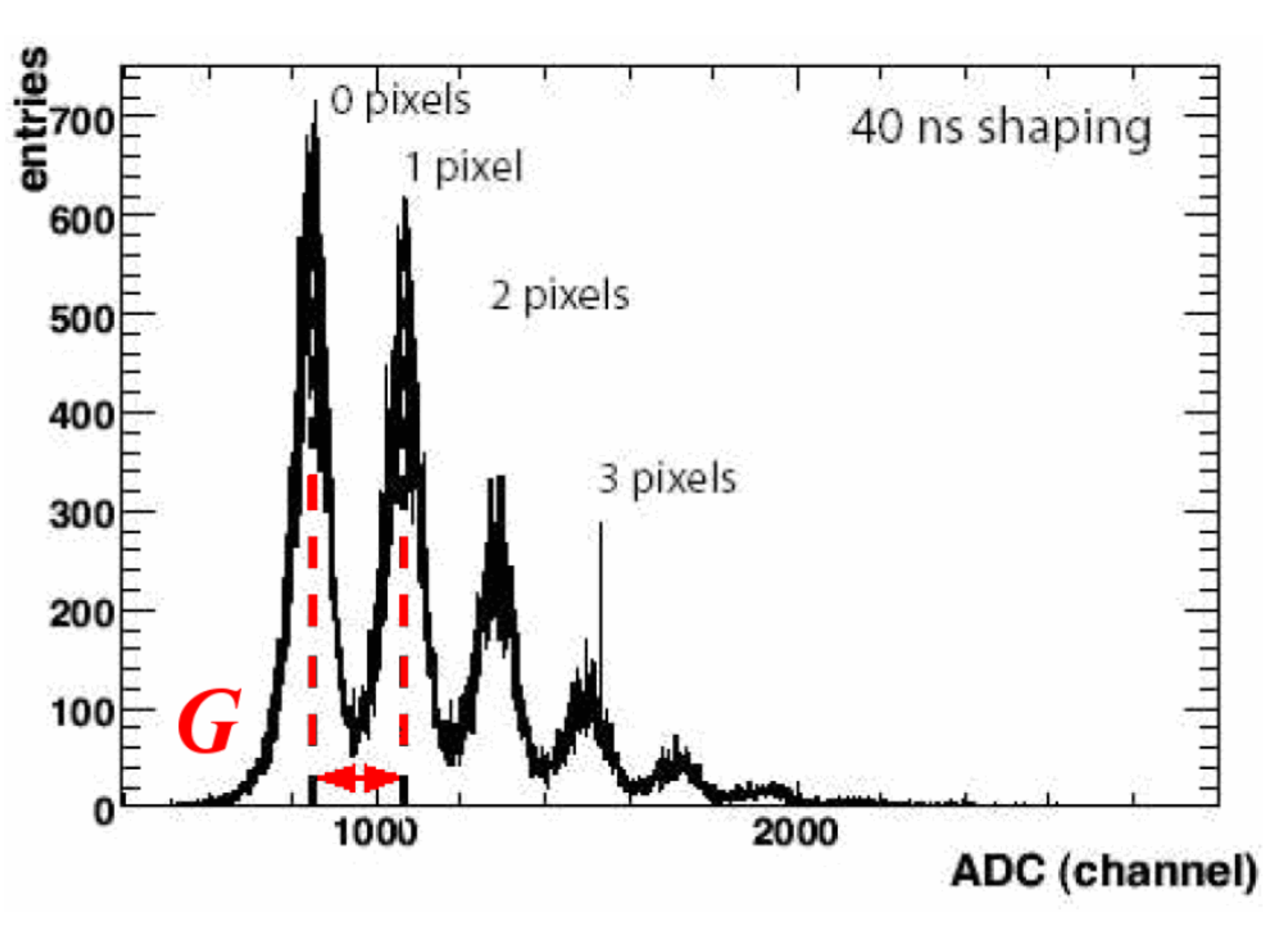}
\caption{Response of a SiPM to low intensity UV LED light coupled into the scintillator tile, recorded with a high gain mode of the front-end electronics. Individual photons can be resolved. The peak to peak distance is used to determine the SiPM gain.}
\label{fig:SinglePhoton}
\end{figure}

The LED system is used to determine and monitor the gain of each SiPM. With low intensity light, only a few photons are detected per pulse. Using a high gain setting of the front-end electronics, the number of pixels fired in the SiPM, corresponding to the number of detected photons, can be resolved. Figure \ref{fig:SinglePhoton} shows an example spectrum recorded in this setting. As illustrated in the figure, the distance between the resolved peaks is used to determine the gain of the SiPMs. Under normal operating conditions, the gain was typically between 10$^5$ and 10$^6$.

SiPMs show a temperature dependence of the gain, where the gain increases with decreasing temperature. This also leads to a temperature dependence of the recorded signal. The LED calibration technique is also used to monitor gain variations with temperature. It has been demonstrated in the CALICE detector setup at Fermilab that the temperature dependence of the SiPM response can be corrected for by a suitable temperature dependent adjustment of the bias voltage. For the data discussed here such a procedure was not applied, however.

The LED system is also used to study the saturation behavior of the SiPMs, which is caused by the finite number of pixels in a sensor, and thus sets a limit for the output signal of the device. This saturation affects the linearity of the response of the calorimeter in regions with high particle density within the shower, such as the core of electromagnetic sub-showers. This study is possible since the LED system covers the full dynamic range of the SiPM, from a few photons up to the equivalent of 200 minimum-ionizing particles crossing a tile, corresponding to several thousand photons. The measured response as a function of LED light intensity is used to correct for the saturation effects in the data analysis.

\section{Beam Studies: Muon Data}

The calibration of each individual tile is performed {\it in situ} with a wide muon beam that illuminates the whole calorimeter. This allows the determination of the ADC value corresponding to the most probable energy loss of a minimum ionizing particle MIP crossing the scintillator tile for each channel in the calorimeter. Typically about 14 pixels fire in the SiPM for the signal of a MIP. From simulations of the passage of 100 GeV muons through 5 mm thick scintillator the most probable energy loss has been determined to be 0.86 MeV, which is used to convert the observed signal from an ADC scale to an energy scale. The most probable value for a MIP is determined for each channel by the fit of a Landau function convolved with a Gaussian to the spectrum recorded with muons. The convolved function is used to account for the energy loss spectrum described by a Landau, as well as for the statistics of photon collection in the cell. Since typically 14 photons are detected, a Gaussian is a good approximation for this distribution. The most probable value extracted from the fit is then used as the calibration constant for the particular channel. 

\begin{figure}
\centering
\includegraphics[width=0.48\textwidth]{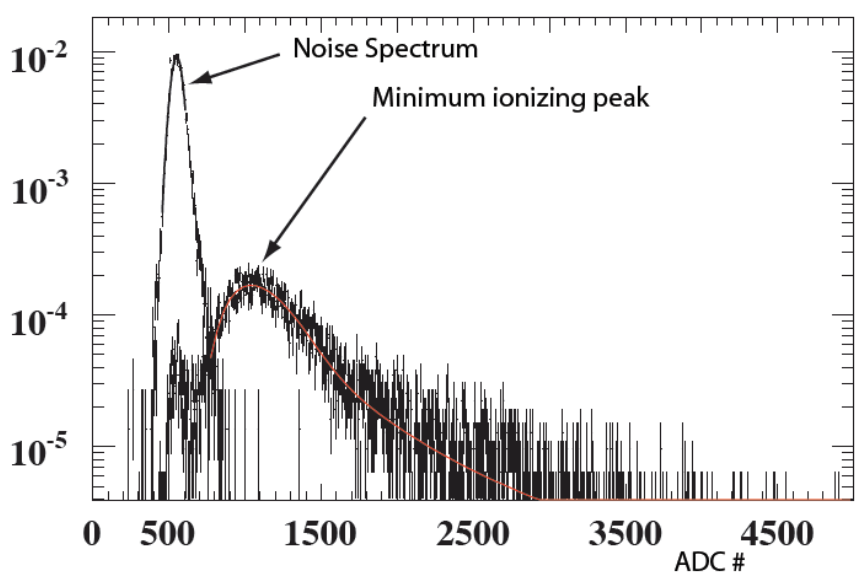}
\caption{Response of a cell to muons. The extraction of the most probable value by a fit is indicated. For reference, the noise spectrum of the same cell is also shown.}
\label{fig:MuonPeak}
\end{figure}

Figure \ref{fig:MuonPeak} shows the response of a typical calorimeter cell to muons, together with a fit used to extract the most probable energy loss. The noise contribution in the signal is small and does not affect the fit result significantly. For illustration purposes, the noise spectrum of that particular cell recorded in a pedestal run is also shown in the figure. For data analysis, a threshold of typically 0.5 MIP is applied to suppress noise contributions. This results in a MIP efficiency of approximately 93\% and a signal to noise ratio of greater than 9.

\begin{figure}
\centering
\includegraphics[width=0.48\textwidth]{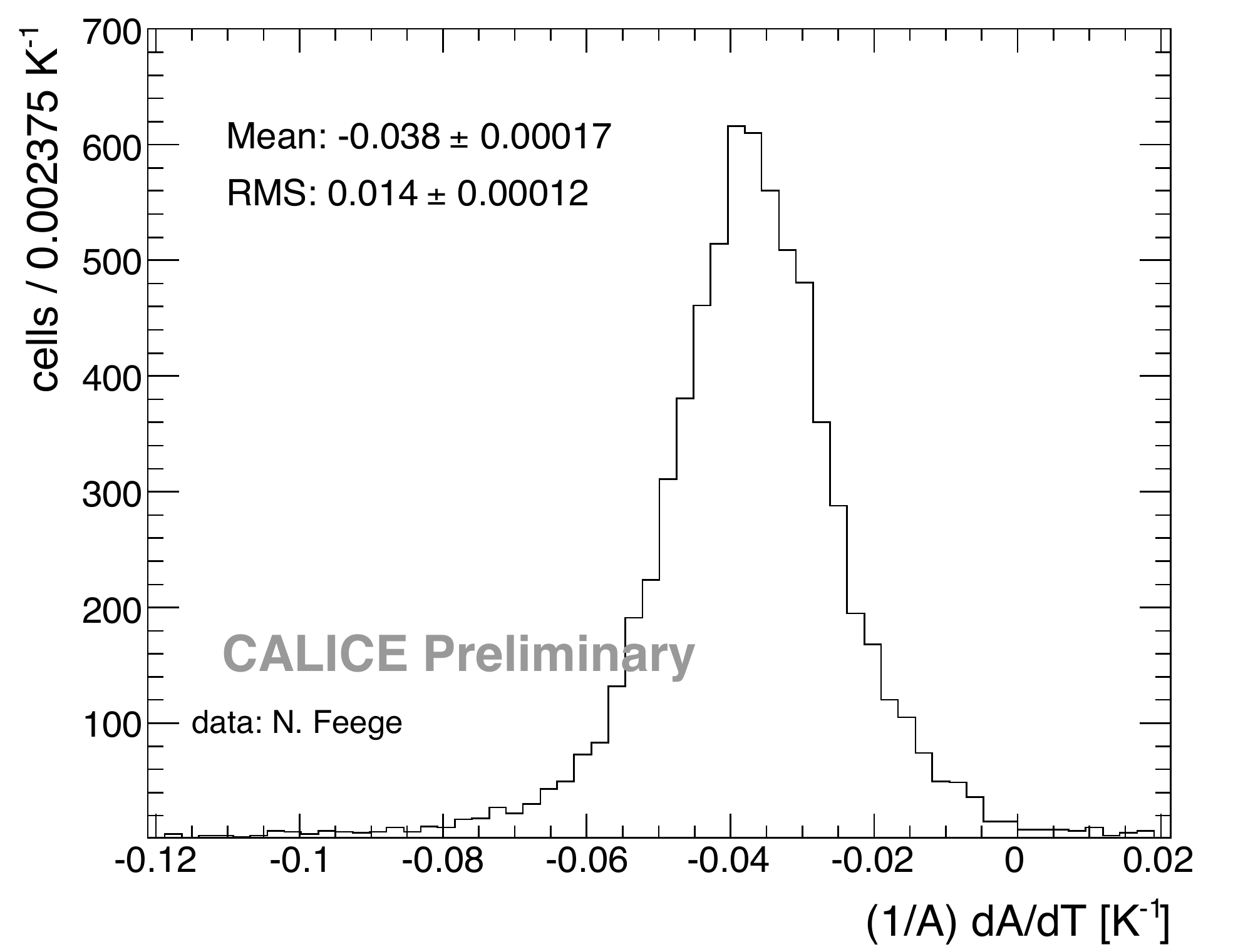}
\caption{Distribution of temperature slopes determined with muon data for approximately 92\% of all cells in the HCAL. The mean slope, giving the mean change of amplitude with temperature for the SiPMs used in the HCAL, is -3.8\% K$^{-1}$ for the full distribution.}
\label{fig:MuonTemp}
\end{figure}

By using muon data taken throughout the 2007 CERN test beam campaign at different detector temperatures, the temperature dependence of the response of the system to particles can be studied. Since the number of muon data sets and thus the possible different temperature values are limited, at most four different temperatures are available per cell. However, since the muon beam illuminated the full calorimeter, measurements are available for most of the cells in the detector. The temperature dependence is extracted from the change of the most probable value for the MIP energy loss as a function of temperature. Figure \ref{fig:MuonTemp} shows the distribution of temperature dependence for $\sim$92\% of all channels in the HCAL. The mean temperature dependence is -3.8\% K$^{-1}$.

\section{Beam Studies: Hadron Data}

In a collider experiment, muons are not readily available in large numbers, so alternatives for the cell-by-cell calibration of the detector are explored. The high granularity of the device is used to identify minimum-ionizing track segments within hadronic showers. Since these segments have properties consistent with muon tracks, they can be used for calibration, if they can be identified cleanly and in sufficient number.

\begin{figure}
\centering
\includegraphics[width=0.48\textwidth]{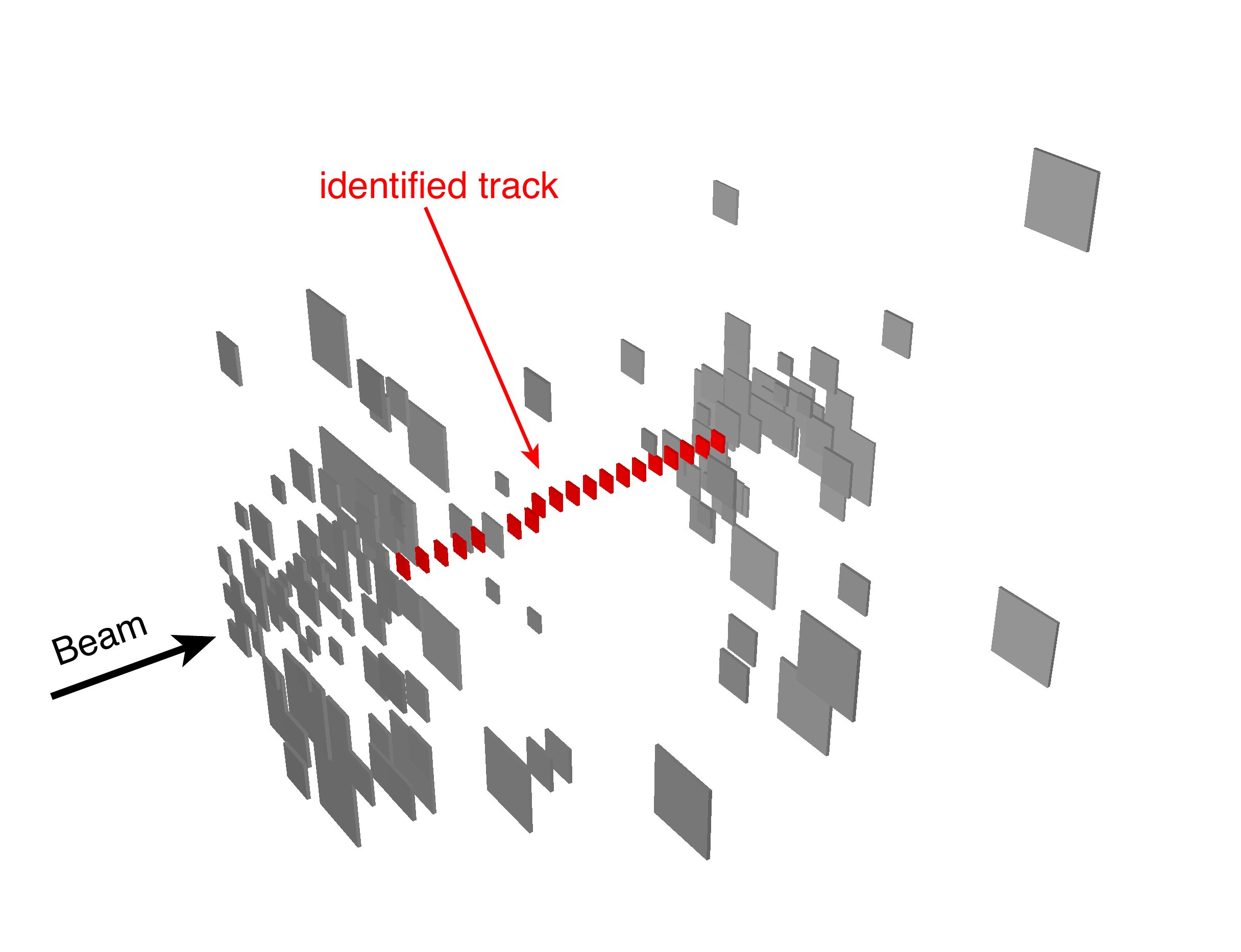}
\caption{Event display of a hadronic shower in the HCAL of a 25 GeV $\pi^-$ with an identified track within the shower. The tiles belonging to the track are highlighted in red, all other tiles with an energy deposit above 0.4 MIP are shown in grey. The size of the cell is the size of the scintillator tile of that particular cell.}
\label{fig:TrackDisplay}
\end{figure}

 Figure \ref{fig:TrackDisplay} shows an example event with one identified track in a data set recorded with a 25 GeV $\pi^-$ beam. This particular event has two high-density sub-showers with presumably high electromagnetic energy content in the HCAL. Those two sub-showers are connected by a minimum-ionizing particle track with a length of approximately 2.3 $\lambda$. The hadronic activity started in the ECAL (not shown in the figure), the most upstream of the CALICE calorimeters in the test beam setup, and propagated into the HCAL. In this particular study, tracks are found from isolated cells which do not have a direct next neighbor with significant energy deposit. The minimum length for a track to be accepted is 6 layers, corresponding to approximately 0.7 $\lambda$. The average length of identified tracks in the 25 GeV data set was 10.5 layers, with a yield of approximately 1.7 tracks per event. 
 
\begin{figure}
\centering
\includegraphics[width=0.48\textwidth]{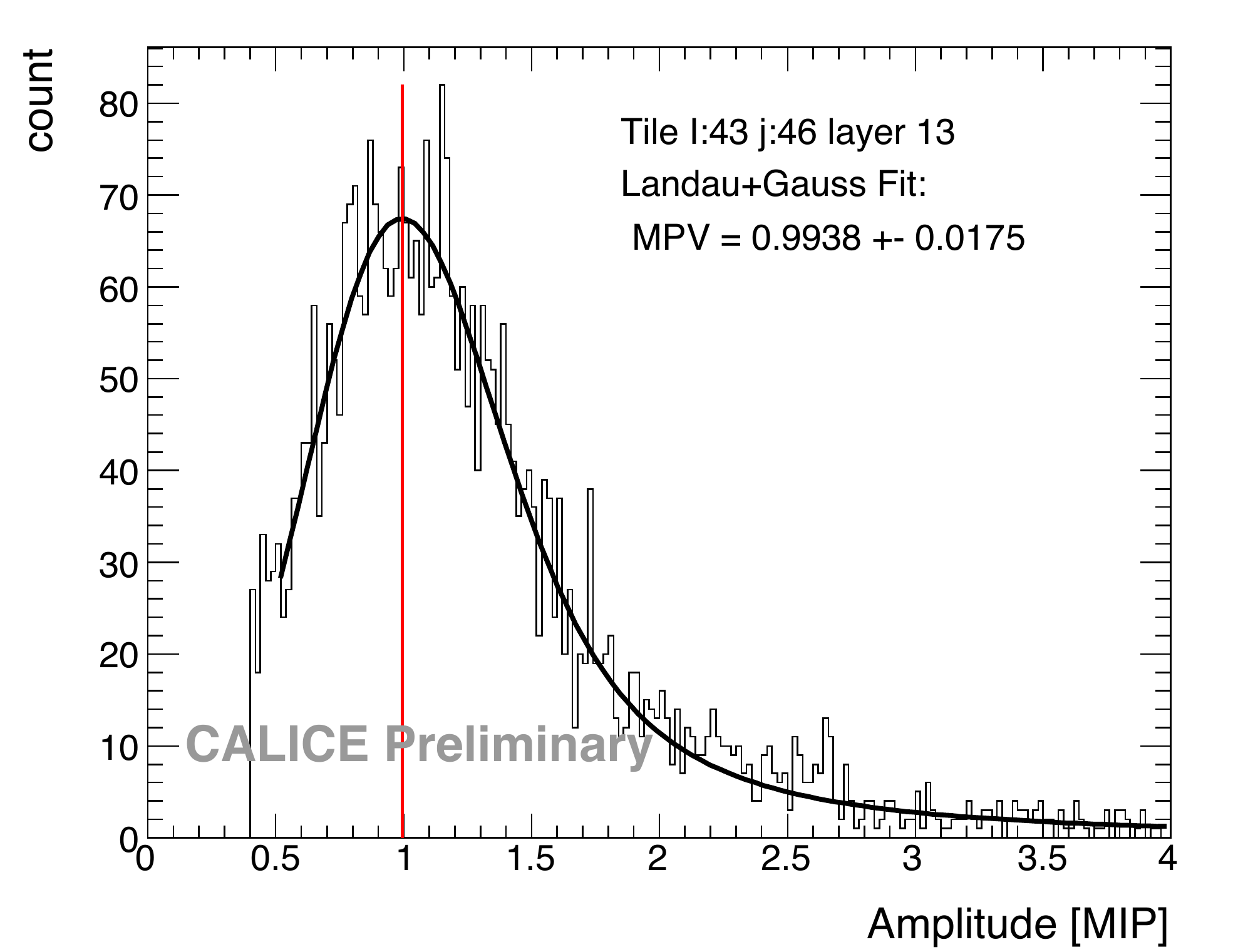}
\caption{Hit amplitude distribution for minimum ionizing track segments identified in hadronic events in a single cell. To extract the most probable value, the distribution is fitted with a Landau convolved with a Gaussian.}
\label{fig:HitDist}
\end{figure}
 
 For each cell that is crossed by an identified track segment, the amplitude is recorded for further study. Figure \ref{fig:HitDist} shows the amplitude distribution for a typical cell. A minimum cut of 0.4 MIP is placed on the amplitude to reject noise on the one hand and to still obtain a significant number of entries below the expected distribution maximum on the other hand. This distribution is fitted with a Landau convolved with a Gaussian to extract the most probable value.

\begin{figure}
\centering
\includegraphics[width=0.48\textwidth]{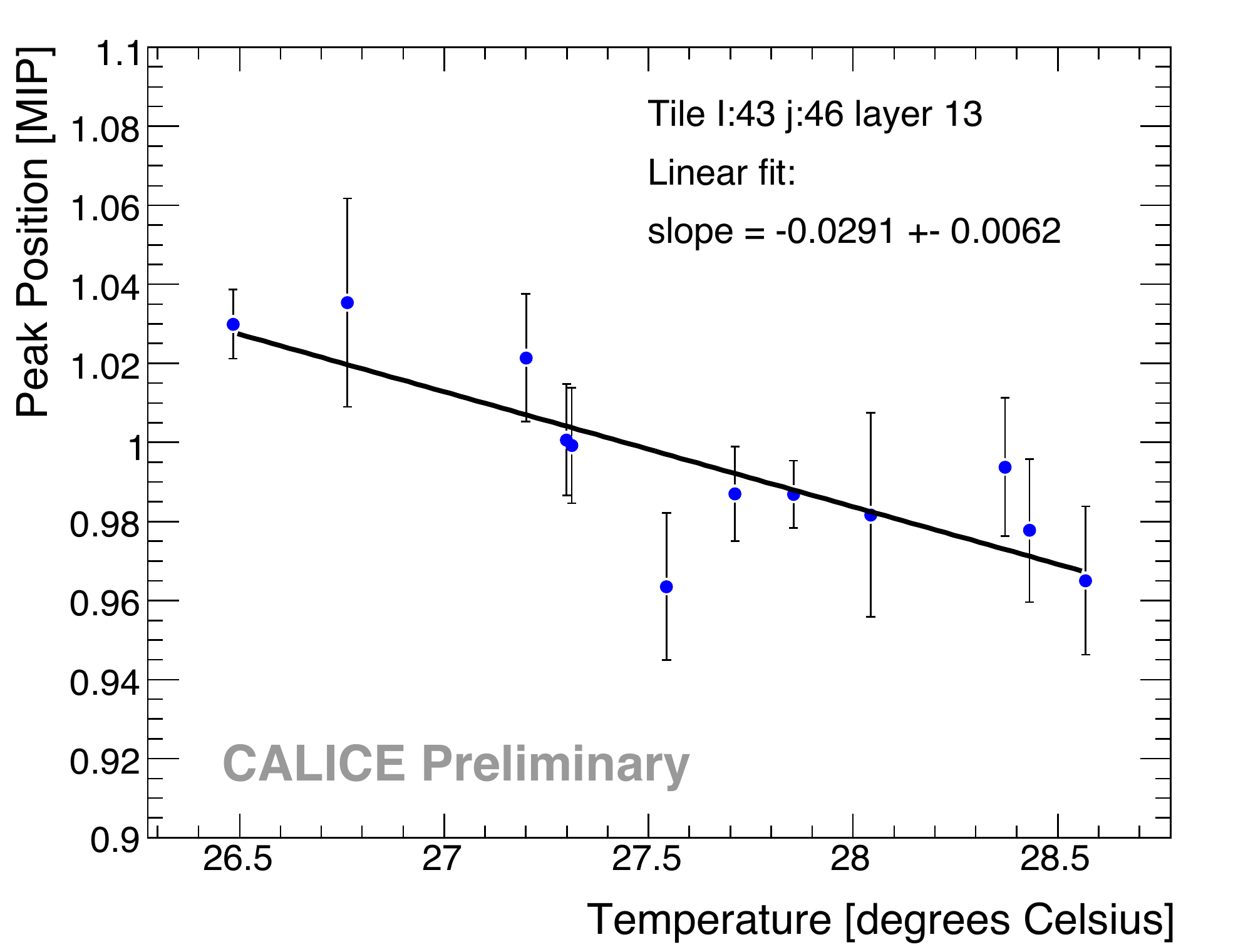}
\caption{Temperature dependence of the most probable amplitude of a single cell for minimum-ionizing tracks in hadron data. In this particular case, a slope of -2.9\% K$^{-1}$ is observed.}
\label{fig:TempSlope}
\end{figure}

The CALICE experiment recorded a large data sample during the 2007 test beam campaign at CERN, which also spanned a large range in ambient temperature. As such, the hadron data is well suited to study the temperature dependent response of the calorimeter. For a subset of 19 hadron runs, the most probable value of the energy loss of track segments in each cell was studied, provided more than 2000 tracks crossed a given cell in a particular run. This requirement restricts the study to the center of the calorimeter, since the number of tracks is highest near the beam axis. Figure \ref{fig:TempSlope} shows the evolution of the most probable amplitude for one cell with temperature, extracted from the hadronic data. For 12 out of the 19 considered runs sufficient statistics for a measurement were obtained. The temperature of the cell was determined using the temperature sensor closest to the cell position in the layer.

\begin{figure}
\centering
\includegraphics[width=0.48\textwidth]{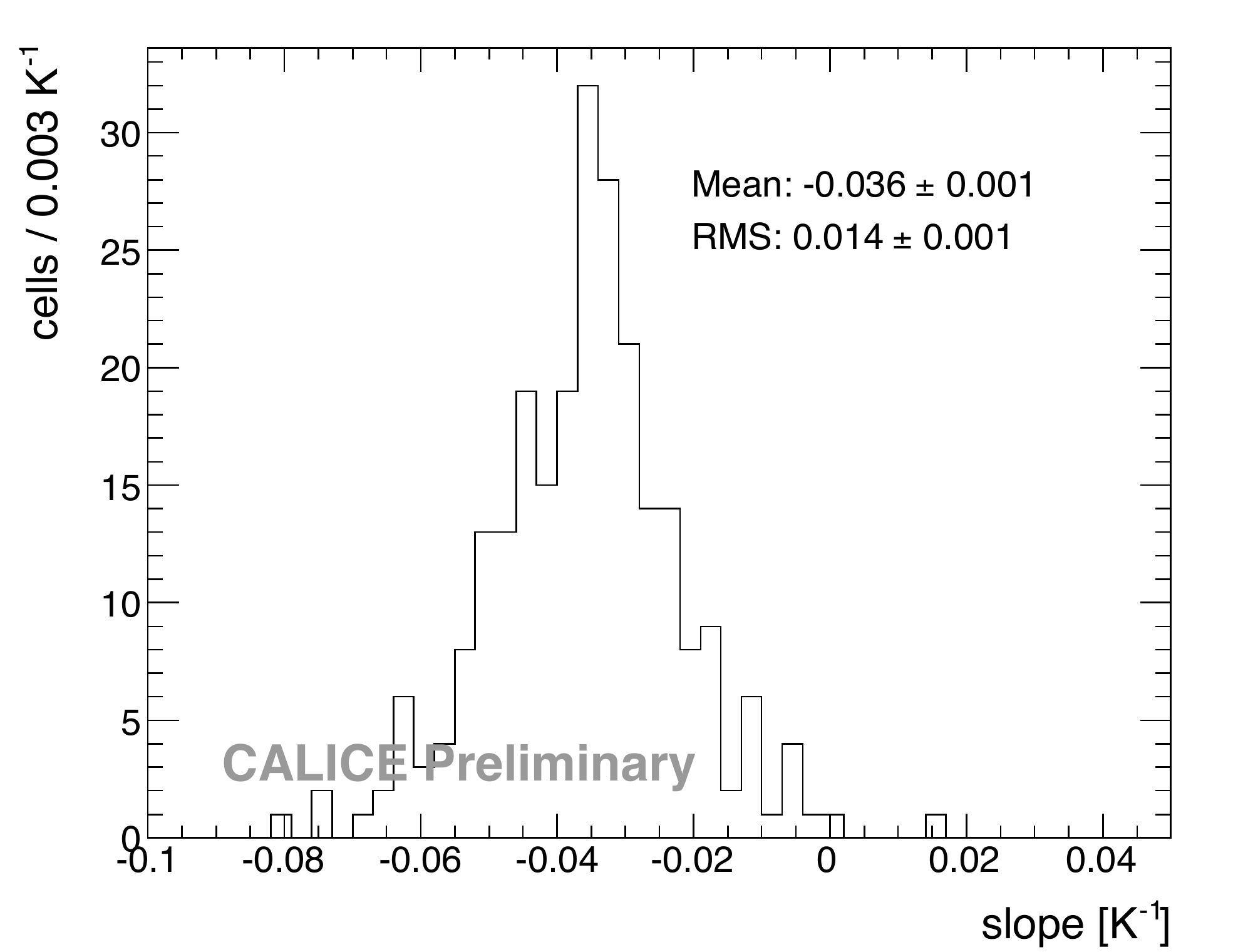}
\caption{Temperature dependence of the response for 250 cells extracted from hadron data. The mean slope, giving the mean change of amplitude with temperature for the SiPMs in the HCAL, is -3.6\% K$^{-1}$.}
\label{fig:SlopeDist}
\end{figure}

Figure \ref{fig:SlopeDist} shows the distribution of the temperature dependence of 250 cells for which measurements at at least 6 different temperatures were obtained in the hadronic data set considered here. The mean value of this distribution, \mbox{-3.6\% K$^{-1}$}, as well as the width, given by the RMS of the distribution, is consistent with the large sample studies using muon data shown in Figure \ref{fig:MuonTemp}. These two measurements are complementary since they were obtained with independent data sets, and while the muon data only spans a very limited number of temperature points, the hadron data covers an extended temperature range, at the expense of the number of studied cells.

\section{Conclusions}

The CALICE collaboration has constructed a highly granular analog hadronic calorimeter prototype based on scintillator tiles with SiPM readout sandwiched between steel absorber plates. This detector has been tested extensively in particle beams. For calibration and monitoring a complex UV LED system, which allows few photon studies in the SiPMs as well as high intensity measurements to explore saturation phenomena, is used. The cell-by-cell calibration is performed using muon data. The high granularity of the device also allows the reconstruction of minimum-ionizing track segments in hadronic showers, which are of sufficient quality to be used for calibration and detector studies in stead of muons. As a first proof of principle measurement, the temperature dependence of the detector response has been studied in both muon and hadron data. The good consistency between the two independent methods demonstrate that the track segments identified in hadronic showers can be used for detailed detector investigations, and might also be a good calibration tool in an ILC detector, allowing the calorimeter to be calibrated with regular data.

\bibliographystyle{IEEEtran.bst}
\bibliography{CALICE}

\end{document}